\documentclass{article}
\usepackage{spconf,amsmath,graphicx}
\usepackage{amsmath,amssymb}
\usepackage{multirow}
\usepackage{graphicx}
\usepackage{epstopdf}
\usepackage{hyperref}
\usepackage{soul}
\usepackage{color,soul}
\usepackage{tabularx}
\usepackage{cite}
\usepackage{subcaption}
\usepackage[english]{babel}
\usepackage{enumitem}
\usepackage[table,xcdraw]{xcolor}
\usepackage{pifont}
\newcommand{\cmark}{\ding{51}}%
\newcommand{\xmark}{\ding{55}}%
\usepackage{makecell}
\usepackage{eqparbox}
\usepackage{booktabs}
\usepackage{multirow}

\newdimen{\algindent}
\usepackage{eso-pic}
\usepackage[linesnumbered,ruled,vlined]{algorithm2e}

\usepackage{bm}

\SetCommentSty{mycommfont}

\SetKwInput{KwInput}{Input}                
\SetKwInput{KwOutput}{Output}              

\title{Synthetic speech detection using meta-learning with prototypical loss}
\name{\begin{tabular}{c}Monisankha~Pal,
        Aditya~Raikar,
        Ashish Panda,
        Sunil Kumar Kopparapu\end{tabular}} 
\address{
TCS Research -- Mumbai, India.} 

\begin{document}
\ninept
\maketitle
\begin{abstract}
Recent works on speech spoofing countermeasures still lack generalization ability to unseen spoofing attacks. This is one of the key issues of ASVspoof challenges especially with the rapid development of diverse and high-quality spoofing algorithms. In this work, we address the generalizability of spoofing detection by proposing prototypical loss under the meta-learning paradigm to mimic the
{\em unseen} 
test scenario during 
training. 
Prototypical loss with metric-learning objectives can learn the embedding space directly and emerges as a strong alternative to prevailing classification loss functions. We propose an anti-spoofing system based on squeeze-excitation Residual network (SE-ResNet) architecture with prototypical loss. We demonstrate that the proposed single system without any data augmentation can achieve competitive performance to the recent best anti-spoofing systems on ASVspoof 2019 logical access (LA) task. Furthermore, the proposed system with data augmentation outperforms the ASVspoof 2021 challenge best baseline both in the progress and evaluation phase of the LA task. On ASVspoof 2019 and 2021 evaluation set LA scenario, we attain a relative 68.4\% and 3.6\% improvement in min-tDCF compared to the challenge 
best baselines, respectively.
\end{abstract}

\section{Introduction}
\label{sec:intro}
\vspace{-5pt}
 Spoofing countermeasure (CM) is needed to make automatic speaker verification (ASV) systems robust against malicious spoofing attacks. In this context, to foster the progress of ASV spoofing CM research, biannual \emph{ASVspoof challenge} series have been organized \cite{kinnunen2017asvspoof, todisco2019asvspoof}. In ASVspoof 2019 challenge logical access (LA) task,
synthetic speech spoofing attacks were produced by voice conversion (VC) and text-to-speech (TTS).
In the current 2021 edition of the same, speech is transmitted across telephone and VoIP networks with several codec and transmission channel effects \cite{yamagishi2021asvspoof}. In recent works on \emph{synthetic speech detection (SSD)}, efforts have been dedicated to develop either novel front-end features or effective classifiers at the back-end. Within this paradigm, recent works on front-end features include cochlear filter cepstral coefficients \cite{patel2015combining}, inverted speech frequency cepstral coefficients \cite{paul2017spectral}, fundamental frequency variation \cite{pal2018synthetic}, linear frequency cepstral coefficients (LFCC) \cite{sahidullah2015comparison}, Constant-Q cepstral coefficients (CQCC) \cite{todisco2017constant} and 
their combinations. On the classifier front, light convolutional neural network (LCNN) \cite{wang2021comparative}, squeeze-excitation residual network (SE-ResNet) \cite{li2021replay}, Res2Net \cite{li2021replay, li2021channel}, end-to-end RawNet2 \cite{tak2021end}, time-domain synthetic speech detection net (TSSDNet) \cite{hua2021towards}, raw spectro-temporal graph attention network (RawGAT-ST) \cite{tak21_asvspoof} have been explored. However, the lack of generalizability remains an issue.
\par
To combat the generalizability issue of spoofing challenges, several data augmentation strategies like on-the-fly frequency augment \cite{chen2020generalization}, signal companding approaches \cite{das2021data} have been utilized on ASVspoof 2019 LA data. Moreover, various codec augmentation methods using codec groups such as VoIP, landline, and cellular facilitates spoofing detection for ASVspoof 2021 LA task \cite{chen21_asvspoof, das21_asvspoof}. The \emph{generalization ability} of the neural network model 
can be improved
by incorporating some specially formulated loss functions. The one-class Softmax (OC-Softmax) loss which sets a compact classification boundary for bonafide class and keeps spoofed class data away by a margin has been successfully applied for spoof detection in \cite{zhang2021one}. In contrast to Softmax classification loss to train an embedding extractor, metric-learning aims at learning the embedding space directly based on some metric or distance function. Deep metric-learning related losses like contrastive and triplet losses were introduced for spoofing detection \cite{gomez2020kernel}, however, these methods are sometimes performance-sensitive and computationally expensive. In this context, \emph{prototypical loss} \cite{snell2017prototypical} to train a protonet under the \emph{meta-learning} paradigm can learn over a variety of spoofing attacks and enable rapid generalization to potentially unseen spoofing attacks. It also uses multiple negative-class samples to learn the embedding space better by prototype representation of each class. 
\par
In this paper, we propose an anti-spoofing system based on SE-ResNet34 architecture with the prototypical loss for both the ASVspoof 2019 and 2021 LA tasks. We consider data augmentation based on known codecs (mentioned in ASVspoof 2021 evaluation plan) such as G711 (alaw) and G722, and random pitch modification and room reverberation for the ASVspoof 2021 evaluation. Finally, we perform a score-level fusion among our proposed 
systems to achieve very competitive results on ASVspoof 2021 LA submission. The main contributions of this work are: (a) empirical analysis to find out the best ResNet-based architecture and comparison among several types of existing and proposed loss functions for SSD task (b) first-ever approach of employing meta-learning with prototypical loss for spoofing CM (c) a new scoring mechanism based on template matching between the two class prototypes and test embeddings.  


\vspace{-5pt}
\section{Proposed system}
\label{sec:format}
\vspace{-5pt}

\subsection{Data augmentation}
\label{sec:dataaug}
\vspace{-5pt}
To address the transmission channel and codec variability in the LA data of ASVspoof 2021, we utilize two known (alaw and G722) codec augmentations. The alaw\footnote{\url{https://www.itu.int/rec/T-REC-G.711}} is a narrow-band landline audio codec. The compression algorithm is based on the ITU-T standard used in Europe and other countries. On the other hand, G722\footnote{\url{https://www.itu.int/rec/T-REC-G.722}} is a wide-band ITU-T standard audio codec mostly used for high-quality applications like VoIP, video conferencing, etc. Since no separate training or development data is provided in ASVspoof 2021 challenge, we apply codec augmentation on all the 
available 
ASVspoof 2019 training and development data. The speech signals are down-sampled, compressed and re-sampled again to $16$ kHz.
We incorporate acoustic simulator\footnote{\url{https://github.com/idiap/acoustic-simulator}} for our codec augmentation.
\par
WavAugment,\footnote{\url{https://github.com/facebookresearch/WavAugment}} an open-source library to implement time-domain augmentation, is used for pitch modification (\emph{pitch}) and room reverberation (\emph{reverb})for ASVspoof 2021 spoofing detection in the LA task. The motivation behind this is that 
some artifacts 
can creep into
synthetic speeches  due to pitch conversion and prediction by VC and TTS algorithms. We randomly select an integer between $+300$ to $-300$ (value calculated by $1/100$ of a tone) and apply that to alter the pitch \cite{kharitonov2021data}. Similarly, for reverb, we sample room-scale between $0$ to $100$ and 
retain 
other parameters 
as mentioned in \cite{kharitonov2021data}. 
As a result,
our training data 
increased 
five fold
after employing the codec and pitch-reverb augmentation. 

\subsection{Embedding extractor}
\label{sec:embedding}
\vspace{-5pt}
In this work, we evaluate and compare the performance of several ResNet architectures, and among them, SE-ResNet34 is used for embedding extraction for all the experiments due to its superior performance over others. ResNet34 adopts `Basic BLK' and each layer follows the same pattern of `Basic BLK' \cite{he2016deep}. The SE block models channel inter-dependencies that give different impact weights to the channels. This can help in exploiting spoofing cues by focusing on channel information \cite{li2021replay}. The two different variants of SE-ResNet34 architectures are adapted from \cite{zhang2021one} and \cite{li2021replay, monteiro2020generalized}; and differs in feature map dimension ([64, 128, 256, 512] and [16, 32, 64, 128]) and pooling mechanism (attentive and global average). These two variants are denoted as SE-ResNet34-atten and SE-ResNet34-avg in this paper respectively.  
\subsection{Meta-learning with prototypical loss}
\label{sec:typestyle}
\vspace{-5pt}
\subsubsection{Motivation}
\vspace{-5pt}
Meta-learning based approaches were introduced to address generalization in few-shot learning \cite{finn2017model}. It is also well-known as learning-to-learn, which learns both on a given task and across tasks. On the other hand, deep metric learning which is one of the approaches to meta-learning learns a metric function in the embedding space \cite{chen2011learning}. We choose prototypical networks or protonet which apply a simpler inductive bias in the form of class prototypes and have been shown to achieve state-of-the-art performance on image classification \cite{snell2017prototypical}, natural language processing \cite{yu2018diverse}, audio classification \cite{Bhosale} and speaker verification \cite{kye2020meta}. Prototypical loss aims to learn the embedding space and correctly classify the unlabeled \emph{query} (test) sample with only a few labeled \emph{support} set (train) samples per class in every \emph{episode} (mini-batch). Therefore, in the pursuit of obtaining generalization over diverse unseen spoofing attacks, prototypical loss with meta-learning might be effective for such a task.

\subsubsection{Episode training}
\vspace{-5pt}
In protonet training, each episode comprises of {$N$ classes} randomly sampled from the total available classes (two for spoofing CM) in training data. In each episode, the labeled set of examples $S = \{S_1, \ldots , S_{{N}}\}$ is the support set is used to construct the class prototypes and the unlabeled query set $Q = \{Q_1, \ldots , Q_{N}\}$ is used to predict classes. In the support set, $S_k = \{\mathbf{x}_i, y_i\}_{i = 1}^{N_S}$ contains $N_S$ 
 labeled samples, where $S_k \subseteq S$, $\mathbf{x}_i$ is a $D$-dimensional feature vector and $y_i \in \{1, \ldots , N\}$. Similarly, within query set $Q_k = \{\mathbf{x}_i, y_i\}_{i = 1}^{N_Q}$, where $N_Q$ represents number of samples for each class in the query set. The protonet learns a non-linear mapping $f_{\boldsymbol{\theta}} : \mathbb{R}^D \rightarrow \mathbb{R}^M$, where $M$ is the dimension of the prototype of each class. The prototype of each class is computed as

\begin{equation}\label{eq1}
    \mathbf{p}_k = \frac{1}{|S_k|} \sum_{(\mathbf{x}_i, y_i) \in S_k} f_{\boldsymbol{\theta}}(\mathbf{x}_i)
\end{equation}
where $\boldsymbol{\theta}$ is the learnable protonet parameters. During training, for every query sample $\{\mathbf{x}_i, y_i\} \in Q$, the posterior probability of class $y_i$ is calculated as
\begin{equation}\label{eq2}
    p_{\boldsymbol{\theta}}(y = y_i|\mathbf{x}_i) = \frac{\textnormal{exp}\big(-d(f_{\boldsymbol{\theta}}(\mathbf{x}_i), \mathbf{p}_{y_i})\big)}{\sum_{k^\prime = 1}^{N}\textnormal{exp}\big(-d(f_{\boldsymbol{\theta}}(\mathbf{x}_i), \mathbf{p}_{k^\prime})\big)}
\end{equation}
In this work, we employ squared Euclidean distance as the distance function $d(.,.)$. Furthermore, prototypical loss within an episode to update the network is calculated as
\begin{equation}\label{eq3}
    L_{\textnormal{PTL}} = \sum_{\{\mathbf{x}_i, y_i \in Q\}} - \textnormal{log} \hspace{2pt} p_{\boldsymbol{\theta}}(y = y_i|\mathbf{x}_i) 
\end{equation}
In spoofing detection, for every episode, the number of classes $N$ is fixed to two and for each class, $N_S$ and $N_Q$ number of samples are randomly taken from the ASVspoof 2019 training set. Algorithm \ref{algo1} shows the episodic training procedure.

\begin{algorithm}[!t]
 
\DontPrintSemicolon
  \KwInput{ $\mathcal{D} = \bigcup_{k=1}^K \mathcal{D}_k$, where $\mathcal{D}_k = \{(\mathbf{x}_i, y_i); y_i = k\}$, Embedding extractor $\boldsymbol{\theta}$, learning rate $\alpha$}
  \KwOutput{Batch training loss $L_{\textnormal{PTL}}$ and model $\boldsymbol{\theta}$}
 
  \For{$k$ in $\{1, \ldots, N\}$}
    {
       $S_k \leftarrow \mathcal{R}(\mathcal{D}_k, N_S)$  \hspace{60pt} \tcc{Supports} 
       $Q_k \leftarrow \mathcal{R}(\mathcal{D}_k \backslash S_k, N_Q)$ \hspace{45pt} \tcc{Queries}
       Calculate prototype of $k$-th class using Eq. (\ref{eq1}) 
    }
  \For{$k$ in $\{1, \ldots, N\}$}
    {
       \For{$(\mathbf{x}_j, y_j)$ in $Q_k$}
         {
            Calculate $L_{\textnormal{PTL}}$ using Eq. (\ref{eq3}) \\
            Optimize $\boldsymbol{\theta}$ with learning rate $\alpha$
         }
    }
\vspace{-5pt}
\caption{Single episode training. $\mathcal{R}(S, N)$ denotes uniform random sampling of N elements from set $S$.}
\label{algo1}
\end{algorithm} 

\subsection{Countermeasure (CM) score computation}
\vspace{-5pt}
After completion of offline training, we use only the trained model to extract embeddings 
from the whole ASVspoof 2019 training set. We compute 
the bonafide and spoof 
class prototypes 
as the mean of corresponding class training embeddings. During the testing, we extract embeddings for each test utterance feature vector $\mathbf{x}_t$ using the same trained model and compute the Euclidean distances from both class prototypes. Our CM score is the difference between these two Euclidean distances and can be written as
\begin{equation}
    \textnormal{Score}_{\textnormal{CM}} = d(f_{\boldsymbol{\theta}}(\mathbf{x}_t), \mathbf{p}_{y=\textnormal{spoof}}) - d(f_{\boldsymbol{\theta}}(\mathbf{x}_t), \mathbf{p}_{y=\textnormal{bonafide}})
\end{equation}

\section{Experimental setup}
\vspace{-5pt}
\label{sec:experimental_setup}

\subsection{Dataset}
\vspace{-5pt}
All experiments are conducted on the ASVspoof 2019 and 2021 challenge LA tracks
\cite{todisco2019asvspoof, yamagishi2021asvspoof}. ASVspoof 2021 doesn't include any separate training and development set and the participants were allowed to use 
the ASVspoof 2019 challenge
partitions. In the LA task, the attacks are the same for both the databases, however, ASVspoof 2021 evaluation data is the degraded version of ASVspoof 2019 evaluation data by various unknown transmission channels. This results in total of $181566$ trials. All the systems are evaluated using challenge-provided minimum tandem detection cost function (min-tDCF) as the primary metric and equal error rate (EER) as the secondary metric \cite{yamagishi2021asvspoof}. The lower the min-tDCF and EER, the better the performance is. 

\vspace{-5pt}
\subsection{Feature extraction}
\vspace{-5pt}
We extract 60-dimensional LFCCs $(\mbox{static}+\Delta+\Delta^2)$ as the front-end features from all the utterances using the MATLAB implementation provided by 
challenge organizers. The frame and hop sizes are $20$ ms and $10$ ms respectively. To form batches, we kept $750$ time frames as the fixed length and for that, we employ repeat padding for short trials similar to \cite{zhang2021one}. For long trials, we randomly choose block of consecutive frames and discard the rest. We also extract log filterbank energy (LFBE) features for some of the experiments. It is worth noting that the features from all the partitions are extracted by considering max frequency present to be $8$ kHz and $4$ kHz for ASVspoof 2019 and 2021 challenge evaluation respectively.


\subsection{Training strategy}
\label{ssec:subhead}
\vspace{-5pt}
We employ our deep ResNet architecture adapted from \cite{zhang2021one}. This deep ResNet34 with attentive pooling is adopted for all the ASVspoof 2019 experiments. The other relatively lighter variant of ResNet34 with global average pooling adapted from \cite{li2021replay} is used for ASVspoof 2021 experiments since it is empirically found to be more effective on ASVspoof 2021 data than the previous one. The block-type considered for ResNet18, ResNet34, ResNet50 and SE-ResNet34 are basic, basic, bottleneck and SE-basic respectively. Speech embeddings with dimension 128 are considered for our proposed system for all the experiments. 
We empirically fix the best combination of the number of episodes, epochs, supports, and queries to be 500, 20, 20, 20 respectively for ASVspoof 2019 evaluation by tuning the accuracy on the development set. For ASVspoof 2021 evaluation, the number of episodes and epochs are increased to 1000 and 100 respectively. 
\par
We train our system with Adam \cite{kingma2014adam} optimizer ($\beta_1 = 0.9$ and $\beta_2 = 0.999$). The batch size is fixed to 64 for all the baselines and for our protonet training episode size is 80. For ASVspoof 2019 evaluation, we set the learning rate initially to 0.0003 with 50\% decay for every 10 epochs, and for ASVspoof 2021 evaluation, we use the learning rate as 0.0005 with 50\% decay for every 15 epochs and that is kept fixed for all the systems. All the models are trained and selected based on maximum accuracy obtained on the development set.

\vspace{-10pt}
\section{Experimental Results}
\label{sssec:subsubhead}
\vspace{-5pt}

\subsection{ASVspoof 2019 LA results}
\vspace{-5pt}
\subsubsection{Experiments on variants of ResNet}
\vspace{-5pt}
We compare the effectiveness of several variants of ResNet-based architectures under prototypical training for spoofing detection. The experimental results are shown in Table \ref{table1}. 
By comparing all the ResNet-types, it is observed that ResNet34 outperforms both ResNet18 and ResNet50, which indicates the bigger model with increasing depth may lead to overfitting. Therefore, we choose ResNet34 and integrate it with SE-block for further improvement. Among two variants of SE-ResNet34, it is seen that SE-ResNet34-atten is better than SE-ResNet34-avg on ASVspoof 2019. The SE block reduces the min-tDCF and EER to 0.0667 and 2.71\% respectively. Finally, we choose the SE-ResNet34-atten model as the best architecture for the following experiments on ASVspoof 2019 data. 

\begin{table}[!t]
\vspace{-5pt}
\caption{Results on the evaluation set of ASVspoof 2019 LA task with different ResNet architectures. LFCC feature front-end and prototypical loss are used to train the models.}
\vspace{-8pt}
\centering
\setlength{\tabcolsep}{4pt}
\begin{tabular}{ccc}
\Xhline{2.5\arrayrulewidth}
System             & min-tDCF & EER (\%)  \\ \Xhline{2.5\arrayrulewidth}
ResNet18          & 0.0880   & 3.52 \\
ResNet34          & 0.0822   & 3.06 \\
ResNet50          & 0.0961   & 4.54 \\
SE-ResNet34-atten & \textbf{0.0667}   & \textbf{2.71} \\
SE-ResNet34-avg   & 0.0980   & 4.80 \\ \Xhline{2.5\arrayrulewidth}
\end{tabular} \label{table1}
\vspace{-8pt}
\end{table}

\begin{table}[!t]
\caption{Results on the evaluation set of ASVspoof 2019 LA scenario using proposed and various other existing loss functions for spoofing countermeasure.}
\vspace{-8pt}
\centering
\setlength{\tabcolsep}{3pt}
\begin{tabular}{cccc}
\Xhline{2.5\arrayrulewidth}
System                       & Loss          & min-tDCF & EER (\%)  \\ \Xhline{2.5\arrayrulewidth}
\multirow{4}{*}{LFCC-SE-ResNet34-atten} & Softmax       & 0.0758   & 3.44 \\
                             & AM-Softmax    & 0.0807   & 3.11 \\
                             & OC-Softmax    & \textbf{0.0486}   & 2.00 \\
                             & Contrastive   & 0.0808   & 3.03 \\
                             & Prototypical   & 0.0667   & 2.71 \\ \hline
\multirow{2}{*}{Spec-LCNN \cite{gomez2020kernel}}  & Contrastive   & 0.1044   & 5.64 \\ 
                             & Triplet       &  0.1001  & 5.15    \\ \hline \hline
LFCC-GMM \cite{todisco2019asvspoof} & -- & 0.2110 & 8.09 \\
LFCC-SE-Res2Net50  \cite{li2021replay} & Cross-entropy & 0.0680 & 2.86 \\
LFCC-ResNet18 \cite{zhang2021one} & OC-Softmax &  0.0590 & 2.19 \\
LFCC-LCNN-LSTM-sum \cite{wang2021comparative}  & P2SGrad     & 0.0524   & 1.92 \\
CQT-MCG-Res2Net50 \cite{li2021channel}         & Cross-entropy & 0.0520   & \textbf{1.78} \\ \Xhline{2.5\arrayrulewidth}
\end{tabular} \label{table2}
\vspace{-10pt}
\end{table}

\begin{table*}[!t]
\vspace{-15pt}
\caption{Details and results of ASVspoof 2021 challenge baselines and our proposed system on progress and evaluation phase of LA task. The absence is denoted as ``–'' in the table.}
\vspace{-8pt}
\centering
\setlength{\tabcolsep}{4pt}
\begin{tabular}{cccccccccc}
\Xhline{2.5\arrayrulewidth}
\multirow{2}{*}{System ID} & \multirow{2}{*}{Feature} & \multicolumn{2}{c}{Augmentation}          & \multirow{2}{*}{Model}           & \multirow{2}{*}{Loss}         & \multicolumn{2}{c}{Progress}    & \multicolumn{2}{c}{Evaluation}  \\ \cline{3-4} \cline{7-8} \cline{9-10}
                           &                          & Codec               & Pitch-reverb        &                                  &                               & min-tDCF        & EER (\%)           & min-tDCF        & EER (\%)           \\  \Xhline{2.5\arrayrulewidth}
1                          & CQCC                     & \multirow{4}{*}{--} & \multirow{4}{*}{--} & GMM             & \multirow{2}{*}{--}           & 0.4948          & 15.80         & 0.4974          & 15.62         \\
2                          & LFCC    &                     &                     &                GMM                  &                               & 0.5836          & 21.13         & 0.5758          & 19.30         \\
3                          &  LFCC                        &                     &                     & LCNN                             & P2SGrad                    & 0.3152          & \textbf{\color{red}8.90}          & 0.3445          & 9.26          \\
4                          &    --                      &                     &                     & RawNet2                          & Softmax                            & 0.4152          & 9.49          & 0.4257          & 9.50          \\ \hline \hline
5                          & \multirow{2}{*}{LFCC}    & --                  & --                  & \multirow{2}{*}{SE-ResNet34-avg} & Prototypical                  & 0.3329          & 10.93         & 0.3545          & 10.39         \\
6                          &                          & --                  & --                  &                                  & OC-Softmax                    & 0.3290          & 10.13         & 0.3510          & 9.81          \\ \hline \hline
7                          & LFCC    &     \cmark                &   \xmark                  & \multirow{5}{*}{SE-ResNet34-avg}                & \multirow{5}{*}{Prototypical} & \textbf{0.3058} & \textbf{9.07} & 0.3412          & \textbf{9.21} \\
8                          &    LFCC                      &    \xmark                 &          \cmark           &                                  &                               & 0.3234          & 10.21         & 0.3429          & 10.68         \\
9                          &     LFCC                     &     \cmark                &        \cmark             &                                  &                               & 0.3899          & 12.53         & 0.3871          & 11.73         \\
10                         & LFBE    &         \cmark            &       \xmark              &                                  &                               & 0.3383          & 11.11         & 0.3538          & 11.14         \\
11                         &   LFBE                       &     \xmark                &        \cmark             &                                  &                               & 0.3354          & 11.34         & 0.3521          & 11.14         \\ \hline
12                         & LFCC                     &     \cmark                &        \xmark             &          SE-ResNet34-avg                        & OC-Softmax                    & 0.3259          & 9.43          & \textbf{0.3407} & 9.54          \\ \hline \hline
7+8+10+11                  & \multirow{2}{*}{--}      & \multicolumn{2}{c}{\multirow{2}{*}{--}}   & \multirow{2}{*}{--}              & \multirow{2}{*}{--}           & 0.3096          & 9.24          & 0.3339          & 9.25          \\
7+8+10+11+12               &                          & \multicolumn{2}{c}{}                      &                                  &                               & \textbf{\color{red}0.3021} & 9.01 & \textbf{\color{red}0.3320} & \textbf{\color{red}9.03} \\ \Xhline{2.5\arrayrulewidth}
\end{tabular}  \label{table3}
\vspace{-15pt}
\end{table*}

\vspace{-5pt}
\subsubsection{Effectiveness of meta-learning with prototypical loss}
\label{sec:print}
\vspace{-5pt}
We compare our proposed meta-learning with prototypical loss function training with the conventional binary classification losses such as Softmax, angular-margin Softmax (AM-Softmax), and one-class Softmax (OC-Softmax) \cite{zhang2021one}. The corresponding results are summarized in Table \ref{table2}. These results show that proposed loss has better generalization ability as compared to Softmax and AM-Softmax while detecting unknown spoofing attacks. The proposed loss function is also better than similar metric-learning based contrastive loss \cite{gomez2020kernel} on the same evaluation setting. Moreover, although not directly comparable, we have reported and included the results of LCNN with contrastive and triplet loss from \cite{gomez2020kernel} evaluated on the same dataset. 
The proposed system significantly outperforms the LCNN with contrastive and triplet loss. This corroborates the superiority of prototypical loss with meta-learning over other metric-learning based loss functions except OC-Softmax loss. Furthermore, it is seen from the table that OC-Softmax loss with SE-ResNet34-atten yields the best performance among all the considered loss functions. Therefore, specially-designed OC-Softmax loss which makes the bonafide class boundary compact while making spoofed class data away from it has slightly better generalization ability than our proposed meta-learning based prototypical loss. We also report the results of challenge best baseline and other existing top systems (no fusion, data augmentation) with front-end features. OC-Softmax with SE-ResNet34-atten emerges as the single best system in terms of min-tDCF metric. It can be seen that the proposed system significantly outperforms the challenge best baseline and most of the other loss-function based systems except OC-Softmax.

\vspace{-10pt}
\subsubsection{t-SNE visualization}
\begin{figure}[!t]
\vspace{-1pt}
\begin{subfigure}{.24\textwidth}
  \centering
  \includegraphics[width=\linewidth, height=3cm]{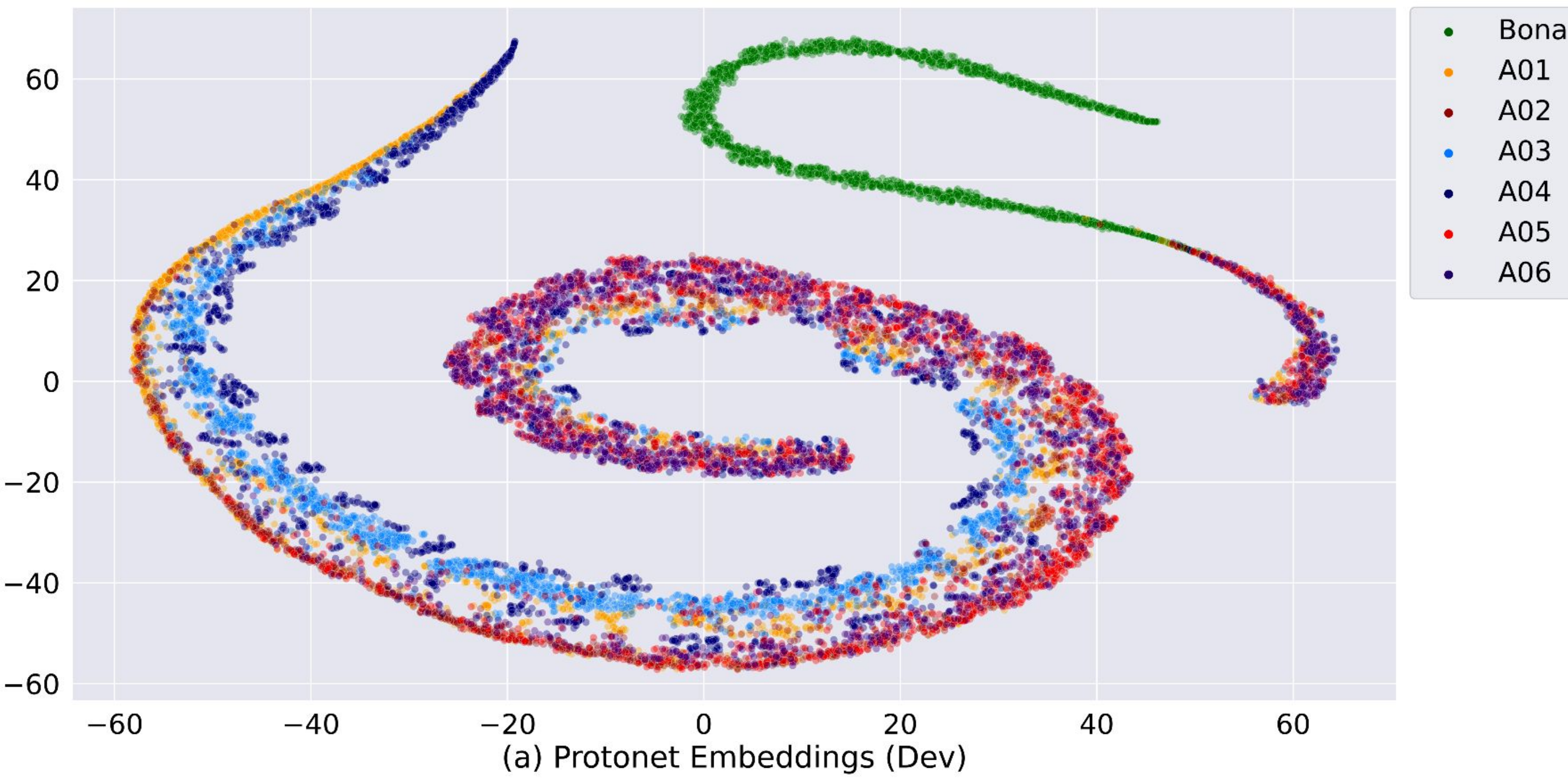} 
  \label{fig:sub-first}
\end{subfigure}
\begin{subfigure}{.24\textwidth}
  \centering
  \includegraphics[width=\linewidth, height=3cm]{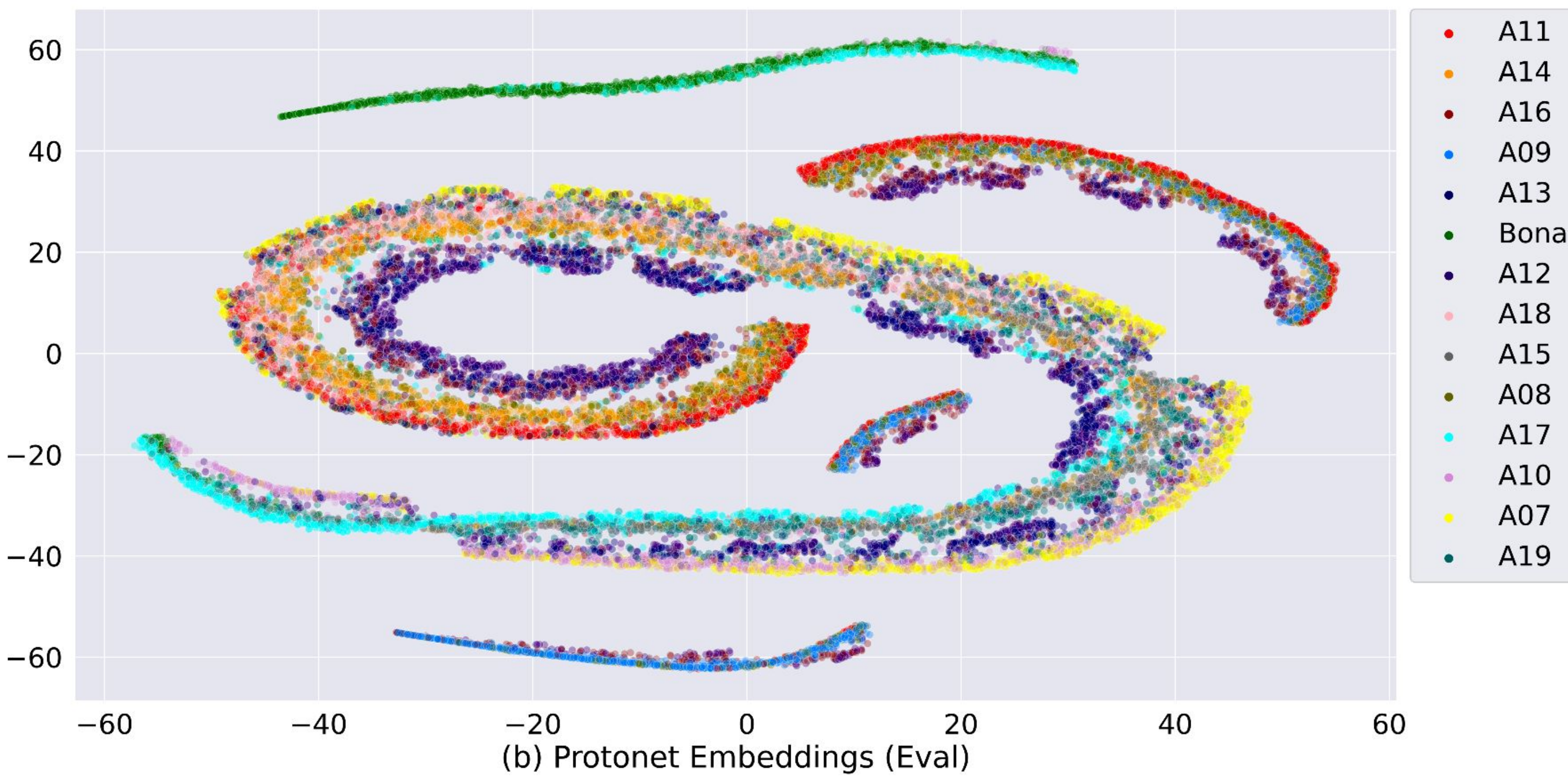} 
  \label{fig:sub-second}
\end{subfigure}
\vspace{-15pt}
\caption{t-SNE visualization of proposed protonet embeddings. Bonafide embeddings are shown in green and all the different attack enbeddings are shown in separate colors.}
\label{fig1}
\vspace{-15pt}
\end{figure}

\vspace{-5pt}
The dimension reduced t-SNE visualization of protonet embeddings (2000 samples for each category) on ASVspoof 2019 development and evaluation set is illustrated in Fig. \ref{fig1} (a) and (b) respectively. The plots are shown in the same x-axis and y-axis scales. It is evident from the figures that bonafide (Bona) samples have similar distributions for both the set and are also well separated from spoofed samples. However, unknown attacks (A07-A19) in the evaluation set have different distributions from the known attacks (A01-A06) of the development set. Nonetheless, some of the A17 and very few A18 attack samples in the evaluation set are overlapping with bonafide speech samples. This suggests our system has good generalization ability to most of the unknown attacks except A17.

\subsection{ASVspoof 2021 LA results}
\vspace{-5pt}
\subsubsection{Performance analysis without data augmentation}
The overall performance analysis in the progress and evaluation phase of the ASVspoof 2021 challenge is illustrated in Table \ref{table3}. Please note that the progress phase is the initial phase of the challenge and the evaluation was done on a subset of the whole database trials. In the evaluation phase results were computed from the remaining trials of the database. 
\par
The first four rows of Table \ref{table3} show the performance of the four challenge baselines on the progress and evaluation phase. It is seen from the table that deep learning based systems perform better than conventional statistical modeling with hand-crafted features. We empirically found that within the proposed setup, the SE-ResNet34-avg model is more effective than the SE-ResNet34-atten model on ASVspoof 2021 data. This suggests that a heavier model with attentive pooling can lead to overfitting issues. Therefore, all the following experiments are based on the SE-ResNet34-avg model. In both the progress and evaluation phase, we obtain significantly better performance using the SE-ResNet34-avg model with prototypical loss as compared to the three challenge baselines and competitive performance to LFCC-LCNN 
Moreover, the proposed system achieves comparable performance to SE-ResNet34-avg with OC-Softmax loss.  

\vspace{-10pt}
\subsubsection{Performance analysis with data augmentation}
\vspace{-5pt}
The effect of our data augmentation approach on the proposed system performance is shown in Table \ref{table3} row 7 to row 12. First, we apply known alaw and G722 codec augmentation and we observe that it significantly reduces the min-tDCF from 0.3329 to 0.3058 and 0.3545 to 0.3412 in progress and evaluation phases respectively. Moreover, we also tried pitch-reverb augmentation separately (row 8) and achieve a substantial boost in performance. However, considering both the codec and pitch-reverb augmentation (row 9) degrades performance. This seems to indicate that applying both these two different kinds of augmentation maybe adding some nuisance information not related to spoofing detection. In this regard, adversarial augmentation strategy similar to \cite{chen21_asvspoof} might help in disentangled speech embedding extraction and further investigation is needed. We also experimented with different front-end features (LFBE) and loss functions (OC-Softmax) and the corresponding results are reported in rows 10 to 12 in the table. Finally, in order to exploit the complementary merits of each developed sub-system, we employ score-level fusion using BOSARIS toolkit \cite{brummer2013bosaris}. The fusion weights are optimized based on ASVspoof 2019 development set. We choose the two best possible combinations, namely 7+8+10+11 and 7+8+10+11+12 and their results are presented in Table \ref{table3}. Finally, we achieve our best system submission in the post-eval phase of the challenge with min-tDCF and EER as 0.3320 and 9.03 respectively. This will rank 18 in the ASVspoof 2021 challenge according to \cite{yamagishi2021asvspoof}, however, as a single system the proposed system is better than some of the top systems \cite{caceres2021biometric, kang2021crim, das21_asvspoof} of the challenge. 

\vspace{-15pt}
\section{Conclusions}
\vspace{-5pt}
In this paper, we introduced meta-learning based prototypical loss to address the generalization issue and developed a robust spoofing detection system for both ASVspoof 2019 and 2021 LA tasks. We investigated the effectiveness of prototypical loss with the SE-ResNet34 model by extensive analysis and comparison against other existing classification and metric-learning based loss functions. The codec and pitch-reverb augmentation substantially improve the spoofing countermeasure performance on ASVspoof 2021 LA task. We achieve promising performance on ASVspoof 2019 evaluation set and a relative 3.6\% and 2.5\% improvement in min-tDCF and EER over the ASVspoof 2021 challenge best LFCC-LCNN baseline in the evaluation phase. In future, it would be worthwhile to explore one-class prototypical loss and other data augmentation strategies in the context of spoofing detection.


\vspace{-10pt}
\footnotesize
\bibliographystyle{IEEEbib}
\bibliography{ref1}
\vspace{-10pt}
\end{document}